\documentclass[12pt]{article}

\usepackage{newtxtext,newtxmath}

\usepackage{graphicx}

\usepackage[letterpaper,margin=1in]{geometry}

\renewenvironment{abstract}
	{\quotation}
	{\endquotation}

\date{}

\makeatletter
\renewcommand{\fnum@figure}{\textbf{Figure \thefigure}}
\renewcommand{\fnum@table}{\textbf{Table \thetable}}
\makeatother

\usepackage{scicite}

\usepackage{url}

\def\scititle{Network science disentangles internal climate variability in global spatial dependence structures}
\title{\bfseries \boldmath \scititle}

\author{
	Arnob Ray$^{1}$, 
	Abhirup Banerjee$^{2}$,
	Rachindra Mawalagedara$^{1}$,
        Auroop R. Ganguly$^{1, 3,4,5\ast}$ \and
	\small$^{1}$Artificial Intelligence for Climate and Sustainability, The Institute for Experiential Artificial Intelligence,\\ \small Northeastern University, Portland, ME 04101, USA.\and
	\small$^{2}$Universität Hamburg, Centrum für Erdsystemforschung und Nachhaltigkeit (CEN), Hamburg, Germany.\and
 	\small$^{3}$Sustainability and Data Sciences Laboratory, Department of Civil and Environmental Engineering, 
    \\ \small Northeastern University, Boston, MA 02115, USA.\and
  	\small$^{4}$The Institute for Experiential Artificial Intelligence, Northeastern University, Boston, MA 02115, USA.\and
   	\small$^{5}$Pacific Northwest National Laboratory (PNNL), Richland, WA 99354, USA.\and
	\small$^\ast$Corresponding author. Email: a.ganguly@northeastern.edu\and
}

\begin{document} 

\maketitle

\begin{abstract} \bfseries \boldmath
A comprehensive characterization of internal climate variability and irreducible uncertainty through initial-condition large ensembles of Earth system models across different spatiotemporal scales remains a significant challenge in climate science. 
In this study, we find significant differences in the spatial connectivity structures of temperature networks across ensemble members, with variations in long-range connections providing a distinguishing feature across the outcomes of initial conditions. Based on this, we introduce a novel quantifier, the `Connectivity Ratio' ($R$), to encapsulate the spatial connectivity structure of each ensemble member by investigating the influence of internal climate variability on the global connectivity patterns in air temperatures. 
$R$ allows us to characterize the variability of spatial dependence structure across the initial condition ensemble members as well as multiple models. Furthermore, we examine changes in spatial connectivity between near-term and long-term projections using $R$, 
which shows a potential shift in climate predictability under anthropogenic influence on a spatial scale.
\end{abstract}

\noindent
\section*{INTRODUCTION}
                                                                    
\par 

{\it Internal climate variability} (ICV), which refers to the natural fluctuations within the climate system independent of external forcings, introduces irreducible uncertainty into climate model projections, leading to substantial variations in regional to global climate patterns, especially over the next few decades \cite{deser2012communication, deser2020certain, upadhyay2023quantifying}. In recent years, with the widespread availability of slightly perturbed initial condition simulations 
 subject to same model and same external forcing, there has been a growing recognition that the presence of ICV complicates the assessment of climate models \cite{jain2023importance, fasullo2020evaluating}, the separation of the forced climate signal from internal variability \cite{deser2016forced, bengtsson2019can, sippel2019uncovering}, the assessment of climate change impacts \cite{schwarzwald2022importance} and the detection of benefits from climate mitigation efforts \cite{tebaldi2013delayed, samset2020delayed}. In response to this recognition, numerous studies have utilized descriptive statistical measures such as ensemble mean, variance, trends, and signal-to-noise ratios and visual comparisons of these measures to characterize the spatial and temporal behavior of ICV \cite{deser2012uncertainty, thompson2015quantifying, mckinnon2018internal, lehner2023origin}. Yet, the methods to fully characterize ICV, particularly at a global scale, remain inadequate. Intrinsic mechanisms within the climate system, such as spatial connectedness, remain largely unexplored as potential ways to characterize ICV, particularly under conditions of irreducible uncertainty, highlighting the need for further investigation in this area.

\par Teleconnections, a natural phenomenon in the climate system, describe the significant association of climatological variables between two long-distant locations, emphasizing spatial dependency patterns that extend from local to global dimensions. These connections demonstrate how changes in one area can impact other regions, forming a climate network within this system. The interplay between large-scale atmospheric circulation patterns and variations in sea surface temperatures connected via ocean-atmosphere energy exchanges drives large-scale climate teleconnection patterns across large geographical areas \cite{yuan2018interconnected, li2021tropical, rezaei2023changes}. Previous studies have investigated different climate modes like El Ni{\~n}o–Southern Oscillation (ENSO), North Atlantic Oscillation (NAO), and their teleconnections under climate change are also investigated under the forced response\ \cite{herein2017theory, haszpra2020investigating, bodai2020forced} based on the large ensemble. As shown in these studies, the spatial connectivity structure of climate variables changes across the globe due to the influence of ICV. Here we propose a novel methodology to characterize ICV  by identifying the differences in spatial connections across the multiple initial-condition ensemble members using a complex systems approach. It offers unique insights into the interdependent nature of climate variability. Consequently, this analysis helps develop a framework for better predicting the global and regional effects of climate change.


\par For this, we employ a climate network (CN) approach, a powerful framework based on complex network theory\ \cite{tsonis2004architecture,donges2009complex}, to analyze the spatio-temporal patterns of global air temperature for future Earth system model (ESM) projections. 
Analyzing the evolution of these networks allows us to understand the changing spatial connectivity structures under the influence of ICV in response to specific external forcing. 
CN utilizes various network quantification measures from complex network theory and provides a concise picture of the underlying mechanism. CN has been widely used to study various climate phenomena, such as, the intricate spatial structure of the global surface temperature\ \cite{donges2009backbone}, heatwave propagation in USA\ \cite{mondal2021complex}, spatio-temporal synchronization of precipitation extremes\ \cite{malik2012analysis,boers2019complex,stolbova2014topology,mondal2020spatiotemporal,agarwal2022complex,banerjee2023spatial}, various teleconnections \cite{steinhaeuser2012multivariate, liu2023teleconnections,gupta2023interconnection,wiedermann2016climate,yamasaki2008climate}, prediction of different climate phenomena\ \cite{stolbova2016tipping,ludescher2021network}, error patterns in numerical weather forecast \cite{gupta2023analysis}, and many other phenomena \cite{fan2021statistical}.
The CN approach is used to assess global teleconnection patterns in historical climate simulations generated by different Earth system models\ \cite{dalelane2023evaluation}. 

\par Our objective is to utilize the intrinsic properties of the CN to articulate the characterization of ICV. We examine the impact of ICV on global connectivity structures formed by pairwise interactions across all grid points in a climate change scenario. Then we use the latent structure of the network efficiently for characterizing ICV. For this, we construct multiple networks using $2$m air temperature data from $72$ realizations of EC-Earth3 ESM\ \cite{doscher2021ec}, and $50$ realizations of MPI-M ESM\ \cite{gutjahr2019max}. We use the near-term future projections ($2020$-$2049$) of the model output for boreal summer under forcing scenario SSP2-4.5 (shared socio-economic pathways). SSP2-4.5 represents intermediate challenges to climate mitigation and adaptation, reflecting future emissions scenarios that are more likely aligned with current greenhouse gas emissions trends \cite{pierre2020global, le2023climate, hausfather2020emissions}. Using various network measures, we study the changes in the network topology in future projections. We use ERA5  air temperature reanalysis data \cite{hersbach2020era5} as benchmark data to compare the network structures obtained from future projections of two ESMs.  
 We propose a new quantifier, {\it Connectivity Ratio} ($R$), to analyze based on the intrinsic structure of different climate networks, which aids in characterizing ICV. The measure $R$ helps us analyze the variation in spatial dependency structures among members of the initial-condition ensemble and various multi-model ensembles. 
 
\par With this $R$, we analyze the change in spatial connectivity structures under climate change by comparing near-term ($2020$-$2049$) and long-term ($2070$-$2099$) future projections. It is widely recognized that due to the substantially longer memory structure of the ocean, there is a strong connection between oceanic and terrestrial climates \cite{shukla1998predictability}. This stronger ocean-to-land connection may increase the long-range predictability of climate over land \cite{krishnamurthy2019predictability}. 
 Our aim is to investigate how these connectivity patterns vary over near-term climate and long-term climate across different initial conditions.
 Our research indicates that, based on our proposed quantifier $R$, the number of initial conditions exhibiting dominant long-range spatial connections decreases in long-term future projections compared to short-term projections at the global scale. This study highlights on a change in predictability \cite{delsole2018predictability}, though the extent varies across different models and spatial scales. Using this $R$, we perform a similar analysis to examine how the connections of specific regions to the global network change across different initial conditions. Our research on understanding connectivity patterns provides a new direction for future studies and has the potential to enhance the understanding of predictability at the intersection of internal climate variability and climate change scenario.


\section*{RESULTS}
\par While most studies emphasize changes in climate patterns grid-wise from local to global scales in response to initial conditions to characterize ICV, we focus on characterizing ICV by examining connectivity patterns due to pairwise interactions among all grid points. This article demonstrates how variations in the spatial connectivity structure of global temperature, influenced by the choice of initial conditions in ESMs, can provide insights into the characterization of ICV.




\begin{figure*}
\centering
\includegraphics[scale=0.65]{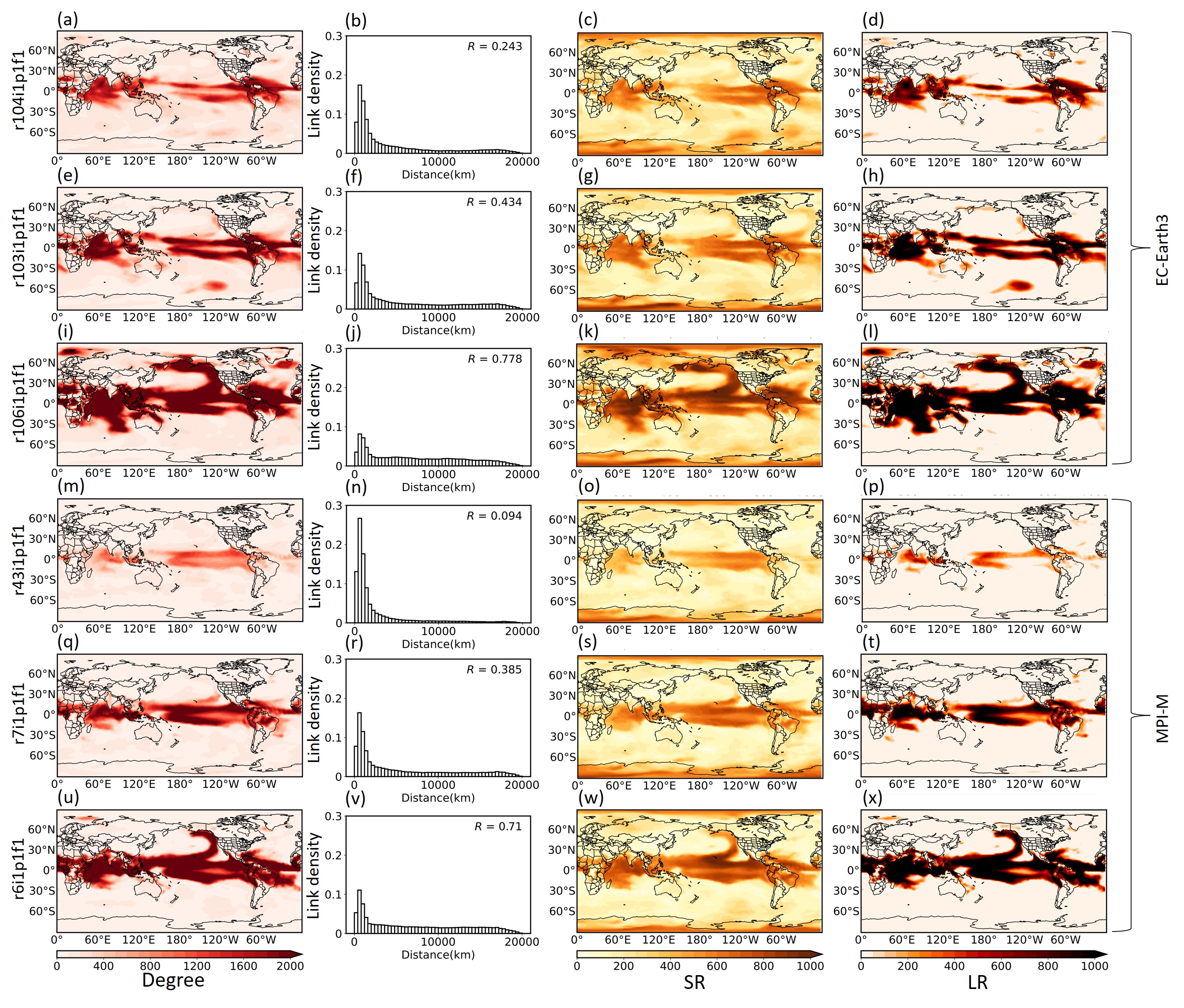}
\caption{{\bf The properties of climate networks generated from the future projections ($2020$-$2049$) of temperature for three different initial conditions for respective EC-Earth3 ESM and MPI-M ESM.} We exhibit the characterizations of climate networks, such as plots of
degree at the global spatial scale in the first column, normalized frequency plots of link lengths along with the value of the connectivity ratio $R$ in the second column, the number of spatial short-range connectivity (`SR') in the third column, and spatial long-range connectivity (`LR') in the fourth column. Here, properties of the network corresponding to different numerical projections of two ESMs show completely different outcomes, exhibiting the influence of climate internal variability on spatial connectivity at the global scale.}
\label{fig1} 
\end{figure*}

\par Here, we choose three datasets of future projections ($2020$-$2049$) obtained from different initial conditions of two ESMs (EC-Earth3 and MPI-M) under the SSP2-4.5 emission scenario so that we can show the variability in the attribute of spatial connections for plausible outcomes of ESMs. After examining the results from all available initial conditions, we use the datasets
from three initial conditions, r104i1p1f1, r103i1p1f1, r106i1p1f1 for EC-Earth3, and r43i1p1f1, r7i1p1f1, r6i1p1f1 for MPI-M, to represent the range of possibilities in the spatial connectivity structure of temperature. The plots of weighted degree (Eq.\ \ref{eq.1}) reveal that the spatial connectivity structures of future projections are qualitatively different depending on the choice of initial conditions of the ESMs, with notable differences in both the number and spatial distribution of links. (first column of Fig.\ \ref{fig1}). We also present the spatial dependency structure of historical data obtained from the ERA5 reanalysis dataset (1990–2019) using a weighted degree plot (Fig.\ \ref{fig3}(a)) provided in the supplementary material. The spatial degree for one initial condition (Figs.\ \ref{fig1}(a) and (m)) of each ESM is qualitatively closer to the spatial connectivity architecture obtained from ERA5 reanalysis data (Fig.\ \ref{fig3}(a)) than other cases. In contrast, the spatial degree for a different initial condition (Figs.\ \ref{fig1}(i) and (u)) differs significantly from the spatial connectivity structure obtained from ERA5 reanalysis data. We observe that high connectivity zones are mainly found in the equatorial regions in all cases, and the possible cause is the presence of the Intertropical Convergence Zone (ITCZ) \cite{wolf2021climate}. Furthermore, we observe strong connectivity with the Pacific Meridional Mode (PMM) \cite{stuecker2018revisiting} for one of the initial conditions (Figs.\ \ref{fig1}(i) and (u)) for both ESMs, which appears less pronounced in the reanalysis data (Fig.\ \ref{fig3}(a)) as well as for the other initial conditions (Figs.\ \ref{fig1}(a) and (e) for EC-Earth3 ESM and Figs.\ \ref{fig1}(m) and (q) for MPI-M ESM). The strong connectivity observed with the PMM in one initial condition but not in others suggests that certain conditions can amplify the PMM’s influence on spatial connectivity in the future. This variability could influence climate patterns in the Pacific, potentially impacting phenomena like the ENSO and its broader climate effects. Overall, these findings illustrate that initial condition differences can lead to diverse outcomes in climate connectivity patterns. This variability suggests a need to consider multiple initial conditions when making climate projections, as single-condition projections may not capture the full range of potential future scenarios. We extend our results for analyzing the connectivity structure from three regions (Ni{\~n}o 3.4, the Amazon rain forest, and the sector of the Indian Ocean) to other locations over the globe in the future (Fig.\ \ref{fig4}) using weighted partial degree (Eq.\ \ref{eq.2}) in Supplementary Materials.

\par Now, based on our previous observation, we are investigating the role of ICV on spatial structures from the perspective of the links of the temperature networks for different cases. The frequency plots of link lengths at different spatial distances also differ according to the intrinsic connectivity patterns of temperature networks (second column of Fig.\ \ref{fig1}). `Fat tail' of the density plots indicates the presence of long-range spatial connections or teleconnections \cite{steinhaeuser2011complex}. A long tail in link density plot suggests the possibility of different teleconnection patterns being present. The thickness of the tail of three plots indicates a large variability of long-range connections in future projections. Moreover, when the tail thickness of density plots increases, the difference between the counts of long-range connection and short-range connection reduces. 
The variation of connections across the initial conditions could be useful to characterize ICV across the initial conditions.    

\par There is no universally accepted rule in the literature for defining the lengths that distinguish between short-range and long-range connections.
Here, we define the {\it short-range connection} and {\it long-range connection} for our investigation. A link is called a short-range connection and a long-range connection if the link length is shorter than $5000$ km and higher than $10000$ km, respectively. We calculate the number of short-range connections (denoted by the `SR') and long-range connections (denoted by the `LR') associated with each node. `SR' of a node indicates the count of links that have lengths shorter than $5000$ km. `LR' of a node is calculated by the sum of the number of links exceeding a length of $10000$ km. Here, the geographical link length, the shortest distance between two locations on a globe, is determined using Haversine's formula\ \cite{van2012heavenly}. For comparison between the total number of long-range connections and short-range connections in a network, we introduce a quantifier, Connectivity Ratio, indicated by $R$ which is defined by calculating the ratio between the sum of `LR' and `SR' over spatial locations (grid points) of the globe. So, the connectivity ratio ($R$) is defined as
\begin{equation}\label{eq.3}
R=\frac{\sum_{i=1}^{m} LR_i}{\sum_{j=1}^{n} SR_j},
\end{equation}
where $m$ and $n$ represent the total number of long-range and short-range connections, respectively, within a climate network. This $R$ is proposed by taking into account the inherent structural properties of the temperature network. The quantifier is effective in capturing the variability of the spatial dependence structure across the initial condition. The larger $R$ values indicate a higher number of long-range connections.  For example, our results show that a $R$ value of $0.7$ always indicates a higher number of long-range connections compared to a $R$ value of $0.4$. Therefore, the value of $R$ can be used to compare the number of long-range connections across ensemble members and capture an intrinsic property of a climate network. If the value of $R$ exceeds $1$,  it indicates that the number of long-range connections surpasses the number of short-range connections.
\par The value of $R$ is determined for each normalized frequency plot of link length (second column of Fig.\ \ref{fig1}). We observe that as the $R$ value increases, the normalized frequency plot of link lengths exhibits a heavy tail, suggesting a higher prevalence of long-range connections. So, the thickness of the tail in the probability density plots of link lengths, which indicates the presence of long-range connections, can be easily interpreted through the values of $R$. As $R$ increases, the tail becomes thicker, signifying a higher likelihood of long-range connections within the network. Conversely, lower values of $R$ correspond to a thinner tail, reflecting a predominance of shorter connections, thereby revealing the relationship between $R$ and the spatial distribution of link lengths. $R$ value easily distinguishes the spatial connectivity pattern generated from one initial condition to another based on the comparison between the number of long-range and short-range connections. 
We generate normalized frequency plots of link lengths and corresponding $R$ values associated with $72$ datasets generated from EC-Earth3 ESM (Fig.\ \ref{fig5}) and $50$ datasets generated from MPI-M ESM (Fig.\ \ref{fig6}) in supplementary material. We can observe qualitative differences ranging from minor to substantial among the individuals for the two ESMs. The variations in short-range connections as well as long-range connections are shown as a result of differences in the internal structure of spatial connectivity based on the choice of initial conditions (third and fourth columns of Fig.\ \ref{fig1}). The quantifier $R$, derived from the spatial connectivity structure, has the potential to be applied from various perspectives.

\begin{figure*}
\centerline{
\includegraphics[scale=0.55]{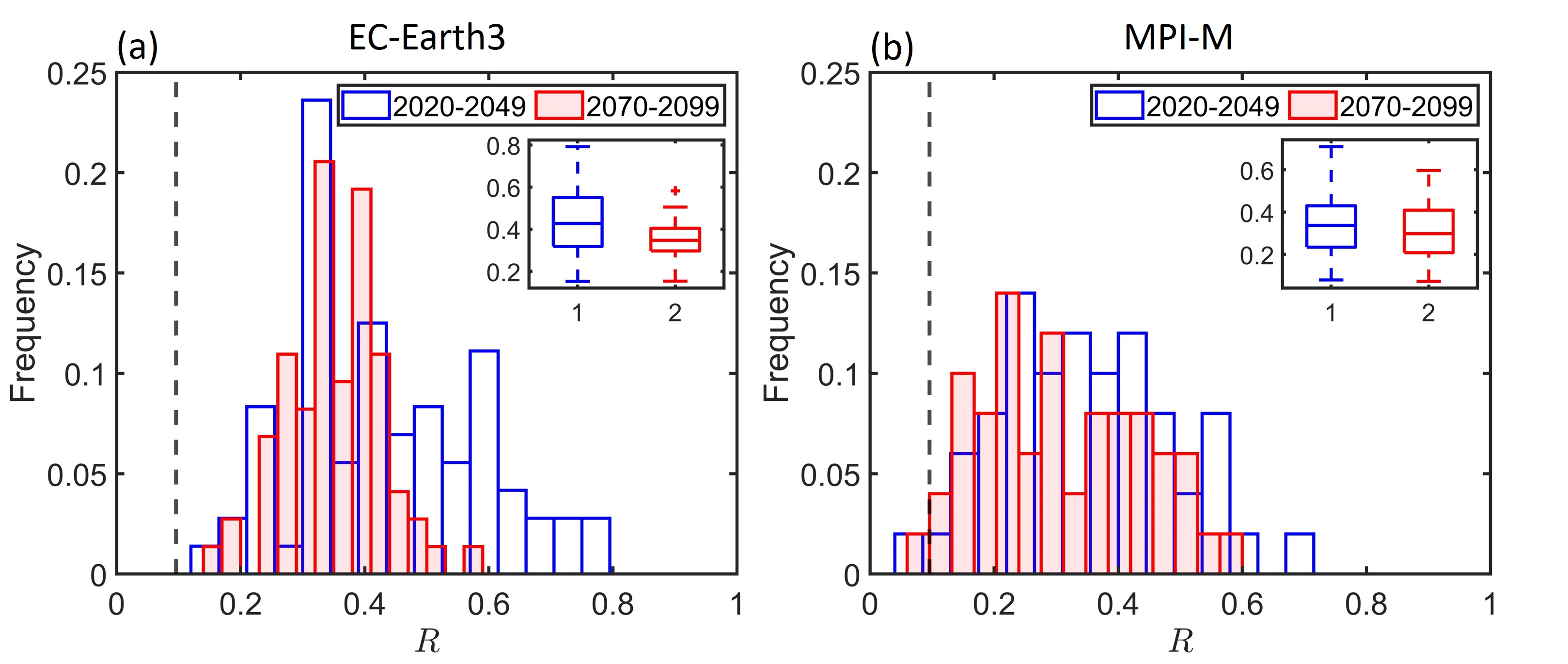}}
\caption{{\bf The histograms of $R$.} We plot normalized frequency plots of $R$ for near-term ($2020$-$2049$) and long-term ($2070$-$2099$) future projection datasets in blue and red for (a) EC-Earth3 ESM and (b) MPI-M ESM, respectively. The variability of the probability density of $R$ significantly decreases for long-term ($2070$-$2099$) future projections compared to short-term ($2020$-$2049$) future projections. We also exhibit box plots for two respective climate scales to support our observed result. The black dashed line indicates the value of $R$ getting from the ERA5 reanalysis dataset.}
\label{fig2} 
\end{figure*}

\begin{figure*}
\centerline{
\includegraphics[scale=0.9]{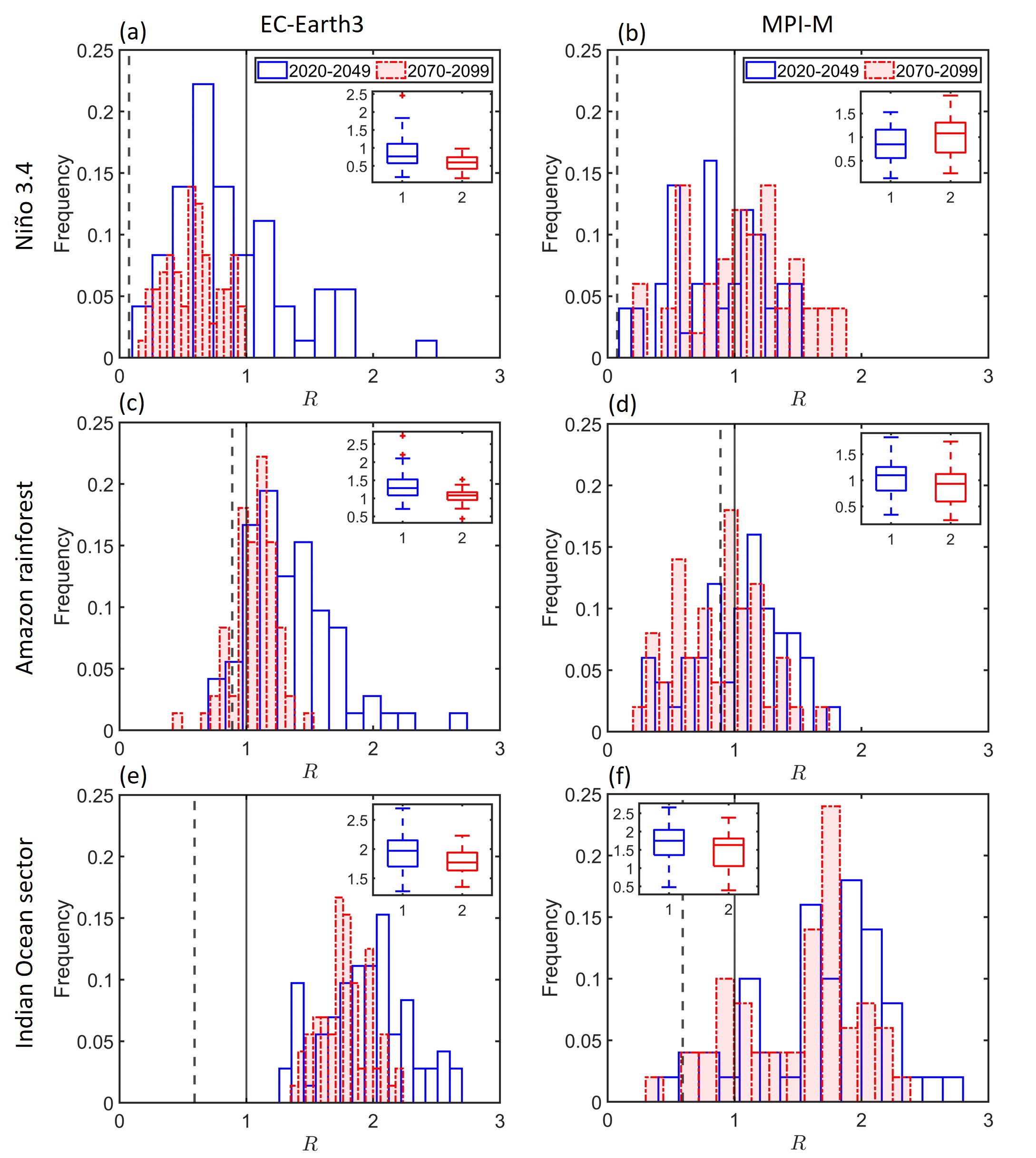}}
\caption{{\bf The histograms of $R$ for three regions}. We plot normalized frequency plots of $R$ for three regions: (a)-(b) Ni{\~n}o 3.4 region, (c)-(d) Amazon rainforest, (e)-(f) sector of Indian Ocean for near-term (2020-
2049) and long-term (2070-2099) future projection datasets in blue and red for EC-Earth3 ESM (first column)
and MPI-M ESM (second column), respectively. We present box plots for the two respective climate scales. The black dashed line represents the $R$ value obtained from the ERA5 reanalysis dataset. This plot demonstrates the variation in long-range connectivity from a particular region to the globe when comparing the two models. The solid black line represents the $R=1$.}
\label{fig7} 
\end{figure*}

\par Tsonis et al.\ \cite{tsonis2008topology} conjectured that temperature predictability, whether high or low, can be reflected in the abundance or scarcity of long-range connections in the temperature network. 
Based on this hypothesis, the $R$ could serve as an indicator of changes in the predictability of various climate models across different spatial scales. This is because $R$ reflects the extent of long-range connections, which are integral to understanding teleconnections and large-scale climate interactions. So, our conjecture is that the variability in frequency of $R$ across different initial conditions in ESMs provides valuable insights into the predictability of climate systems under anthropogenic influences. Here, we will discuss the changes in spatial connectivity structure between near-term future projections ($2020$-$2049$) and long-term future projections ($2070$-$2099$) at the global scale using this quantifier $R$ as an application of initial-condition large ensembles of two ESMs. We consider monthly (June-July-August) 2m air temperature data for $72$ initial conditions of EC-Earth3 ESM and for $50$ initial conditions of MPI-M ESM, over $2020$-$2049$ and $2070$-$2099$ and compute two sets of $R$ values for the near-term and long-term future projections for both ESMs. Normalized frequency plots of $R$ for the respective EC-Earth3 (Fig.\ \ref{fig2}(a)) and MPI-M (Fig.\ \ref{fig2}(b)) ESMs with the $R$ value generated for the ERA5 dataset (indicated by the black dashed line in Figs.\ \ref{fig2}) are shown for capturing the variability of $R$ in terms of large ensembles simulations. The variance in frequency of $R$ is observed to be higher in near-term future projections (blue) compared to long-term future projections (red) across both ESMs. This pattern is further corroborated by box plots (blue for near-term and red for long-term climate scales) for the two datasets (Fig. \ref{fig2}). These box plots\ \cite{mcgill1978variations} demonstrate a reduction in the variability of $R$ values over the long-term climate scale relative to the near-term climate scale for both scenarios. The frequency of initial conditions that display long-range spatial connections reduces for long-term future projections compared to near-term future projections. The dominance of long-range spatial connections, in terms of the number of initial conditions, decreases. Particularly, predictability decreases by the end of the century because most of the initial conditions from the two climate models show a reduction in long-range teleconnections at the global scale. Furthermore, the Kolmogorov-Smirnov (KS) test \cite{massey1951kolmogorov} is employed to evaluate whether the probability density distributions of the two datasets are statistically significant, thereby determining if they originate from the same underlying distribution at the $95\%$ confidence interval. So, the analysis reveals a significant difference between the probability density functions for near-term and long-term future changes in EC-Earth3. In contrast, the probability density functions for MPI-M are statistically similar, highlighting distinct variability patterns between the two models.
A statistically significant change in predictability between the two climate scales is observed for EC-Earth3, while no such change is detected for MPI-M.
It provides qualitative insights into the comparison of near-term climate and long-term climate between two models, utilizing the large ensembles of initial conditions.



\par  Next, the normalized frequency plots of $R$ getting for all considered initial conditions for three regions (Ni{\~n}o 3.4, the Amazon rainforest, and a sector of the Indian Ocean) are showcased to compare the differences in near-term and long-term climate futures between the two ESMs (Fig.\ \ref{fig7}). We include the $R$ value derived from ERA5 (represented by a black dashed line) and the reference line $R=1$ (depicted as a black solid line). Here, the link lengths for a specific region are defined as the geographical distances between points within the region (Fig.\ \ref{fig4}) and all grid points across the globe. For EC-Earth3 ESM (first column of Fig.\ \ref{fig7}), the variances of the frequency plots $R$ are higher for near-term future projections (blue) compared to long-term future projections (red) across all three regions considered, which is in the agreement with global scale findings. Additionally, the differences in link densities are statistically significant at the $95\%$ confidence interval (by performing KS statistical significance test). However, for the MPI-M ESM, the variance of the frequency distribution of $R$ for near-term future projections (in blue) is smaller than that for long-term future projections (in red) for Ni{\~n}o 3.4 (Fig.\ \ref{fig7}(b)), while it is higher for the other two regions: the Amazon rainforest (Fig.\ \ref{fig7}(d)), and a sector of the Indian Ocean (Fig.\ \ref{fig7}(f)). However,  similar to the global scale results, the changes in the behavior of $R$ are not statistically significant for MPI-M at a $95\%$ confidence interval for all three regions. Although the qualitative change in $R$ across initial conditions, reflecting predictability differences between the two future climate scales, is evident in the EC-Earth3 ESM, the change is not significant for the MPI-M ESM for all regions. A comprehensive analysis for each of the three regions is provided in supplementary material.  
\par As demonstrated above, the quantifier $R$ is an effective quantifier in analyzing and comparing the spatial structures of temperature for different ICs within an ensemble. Furthermore, it facilitates comparison between multi-model multi-initial condition ensembles, providing a robust framework for evaluating variability and agreement of spatial connectivity structure.  Consequently, $R$ emerges as a valuable quantifier for model comparison in the context of teleconnection studies.



\section*{DISCUSSION}\label{discuss}
\par We provide a framework for characterizing ICV based on the spatial dependence structure of temperature. We propose a quantifier, connectivity ratio $R$, based on the internal structure of the climate network, i.e. the ratio of the number of long-range and short-range connections. This method underlines the interconnectedness among the 2m air temperature of June-July-August across the globe. Here, the temperature datasets generated for $72$ initial conditions of the EC-Earth3 ESM, and $50$ initial conditions of the MPI-M ESM under the same forcing conditions are considered for our study. 
\\

\par We now have a single number to quantify the behavior of different ensemble members making it possible to characterize ICV across initial conditions. As an example application, $R$ can be used to capture changes in the variability of the internal structure of a climate network over time periods. We have demonstrated this by showing how the probability density plots of $R$ across multiple initial condition ensembles can distinguish between near-term and end-of-century climate networks constructed from climate projections. We further propose that this quantifier, connectivity ratio has the potential to be used for the evaluation of multi-model and multiple initial-condition large ensembles with $R$ values providing a method for direct comparison of different models using ensemble members.


\par In the analysis of spatial connectivity structure, we observe a stronger pattern of ocean-to-land connections with some initial conditions exhibiting a substantially higher number of long-range connections. In such cases, it would be reasonable to infer that the predictability of land climatology would also improve relative to a situation where ocean-to-land connections are weaker. Initial condition simulations with higher predictability could offer valuable insights for climate model development and for advancing our understanding of predictability at climate scales.
Furthermore,  as suggested by Wang et al.\ \cite{wang2024characteristics}, there appears to be a probable relationship between a model’s prediction skills and the network structure. Applying a similar approach to an initial condition ensemble could provide a better understanding of the prediction skills of climate models. Furthermore, the community structure \cite{steinhaeuser2011complex} can be analyzed across initial-condition large ensembles of ESMs to examine how it varies between different initial-condition simulations. This analysis could offer deeper insights into the characterization of ICV, its quantification, and studies on predictability. 
Climate resilience decisions and the policies associated with them, stand to benefit across a broad range of sectors from such insights. For example, the food-energy-water nexus, comprising interdependent systems that rely on stable and predictable climate conditions, can utilize this enhanced understanding of predictability to better manage resources, avoid shortages, and reduce the risk of climate-related disruptions. Improved climate forecasts can also aid in the preservation of ecosystem services, helping to protect biodiversity, manage forests, and conserve freshwater resources.
This insight will assist climate scientists and decision-makers in deriving robust scientific conclusions from large ensembles while considering all sources of potential uncertainty, thereby enabling robust decision-making across a wide range of plausible outcomes.

\clearpage 

%
\bibliography{science_template} 
\bibliographystyle{sciencemag}

\paragraph*{Data and materials availability:}
CMIP6 data is available from \url{https://aims2.llnl.gov/search/cmip6/}. ECMWF Reanalysis ERA5 data is available from \url{https://cds.climate.copernicus.eu/}.




\newpage



\renewcommand{\thefigure}{S\arabic{figure}}
\renewcommand{\thetable}{S\arabic{table}}
\renewcommand{\theequation}{S\arabic{equation}}
\renewcommand{\thepage}{S\arabic{page}}
\setcounter{figure}{0}
\setcounter{table}{0}
\setcounter{equation}{0}
\setcounter{page}{1} 


\begin{center}
\section*{Supplementary Materials for\\ Network science disentangles internal climate variability in global spatial
dependence structures}

Arnob Ray,
Abhirup Banerjee,
Rachindra Mawalagedara,
Auroop Ganguly$^{\ast}$,
\small$^\ast$Corresponding author. Email: a.ganguly@northeastern.edu\\
\end{center}

\subsubsection*{This PDF file includes:}
Materials and Methods\\
Supplementary Text\\
Figures S1, S2, S3 and S4\\
Table S1\\




\subsection*{Materials and Methods}\label{data_method}
\subsubsection*{Data}\label{data}
We utilize the EC-Earth3 ESM \cite{doscher2022ec} and MPI-M ESM \cite{gutjahr2019max} from the CMIP6 data archive, selected for its extensive range of initial condition ensembles. We analyze monthly 2m air temperature (\textit{tas}) data for $72$ and $50$ different initial conditions (ICs) over two periods: 2020-2049 (near-term future projection), and 2070-2099 (long-term future projection) of respective models. Each initial condition is denoted as 'rxi1p1f1', with 
`x' varying based on the realization of the initial state. The simulations for each IC are subjected to the same external forcing while varying the initial conditions. For future projections, we incorporate shared socio-economic pathways (SSP) \cite{meinshausen2019ssp}, specifically SSP2-4.5, to account for climate change scenarios. We also use monthly 2m air temperature (\textit{t2m}) of ERA5 Reanalysis data \cite{hersbach2020era5} for the period 1990-2019 as our reference dataset. To facilitate our analysis, we calculate standardized monthly temperature anomalies over 30 years to mitigate the effects of seasonality. In this study, we focus on the boreal summer months (June-July-August) for our investigation. We regrided all the datasets to $2^\circ \times 2^\circ$ spatial resolution.

\subsubsection*{Methodology}\label{methodology}

\subsubsection*{{\it Climate network construction} }

Climate network analysis provides a robust framework to analyze the spatio-temporal variability of climate variables, which encode the pairwise interactions between
its components\cite{fan2021statistical,donges2009complex,tsonis2004architecture}. In a climate network, each geographical grid point in the spatio-temporal data is considered a node, and an edge is established by computing the statistical
inter-dependency between the time series of the respected grid points.  
Several similarity measures are described in the literature\ \cite{fan2021statistical} that are useful for generating a network based on the characteristics of a time series. 
In our study, we use the Pearson correlation measure for quantifying the statistical interdependencies between pairs of time series\ \cite{donges2009complex, saha2023signatures}.
We conduct the pairwise correlation analysis for all the grid points in our data (N=16200) and construct an $N \times N$ similarity matrix.

\par A connection is determined between two nodes when the $p$-value of the associated correlation is less than $0.05$ (statistically significant of the correlation coefficient) and the absolute correlation coefficient is greater than a predefined threshold, which is defined as $0.5$ \cite{steinhaeuser2011complex, wang2024characteristics}. So, we get the adjacency matrix associated with a CN, which is defined as 

\begin{equation*}
    A_{ij} = 
\begin{cases} 
1 & \text{if}~\{x_i(t)\}~\text{is significantly correlated to}~\{x_{j\ne i}(t)\}\\
0 & \text{otherwise}
\end{cases}
\end{equation*}

where, $\{x_i(t)\}$ is the time series of $i$ the node and $i$, $j$ $\in \{1,2,\cdots,16200\}$. 

\subsubsection*{\it{Network quantification}}

\begin{itemize}
    \item {\bf Degree}: To quantify the relationships between different nodes (geographical grid points) in the network, we use the basic network measure called {\it degree}. This measure computes the number of connections a node $i$ has with other nodes in CN. The weighted degree at node $i$ is defined as 
\begin{equation}\label{eq.1}
k_i=\sum_{j=1}^{N}cos(\lambda_i)A_{ij},
\end{equation}
where $\lambda_i$ is the latitude of $i$-th node. This degree at each node over the globe estimates the number of locations has a strong statistical relation to the time series $\{x_i(t)\}$. We consider the weight of each node by multiplying the cosine of the latitude associated with that location to avoid an inhomogeneous link distribution of the globe \cite{heitzig2012node}.

\item {\bf Partial degree}: For the case study of spatial connectivity structure with respect to a specific region, we use the metric {\it partial degree}\ \ \cite{gupta2023interconnection, jamali2023spatiotemporal, gupta2023analysis} of a node of the CN that calculates the number of links connecting the corresponding node with the nodes within the region. It quantifies the spatial connectivity between the nodes in a certain region $R$ with other nodes in the network. The partial degree of a node $i$ with respect to $R$ is defined as
\begin{equation}\label{eq.2}
k_i^{R}=\sum_{j\in R}cos(\lambda_i)A_{ij}.
\end{equation}
This quantifier provides insights into the connectivity of the spatial relationships between the nodes in a certain region and other locations throughout the world.

\item {\bf Distance between two probability density plots}: To distinguish between two network structures, we use the {\it Jensen-Shannon divergence} \cite{lin1991divergence}, which is used to calculate the difference between two probability density functions. Suppose we have a set of probabilities  $\{p_1, p_2,\cdots,p_n\}$ whose distribution function is $P$ and another set of probabilities $\{q_1, q_2,\cdots,q_n\}$ whose distribution function is $Q$. Then Jensen-Shannon divergence is defined by   
\begin{equation}\label{eq.4}
    D_{JSD}(P||Q) = \frac{D_{KL}(P||M)+D_{KL}(Q||M)}{2} 
\end{equation}
where, $M = \frac{P+Q}{2}$. Now, $D_{KL}(P||Q)$ indicates the Kullback–Leibler divergence \cite{kullback1951information} which is defined by 
\begin{equation*}
    D_{KL}(P||Q) = \sum_{i=1}^n p_i~log\Bigg(\frac{p_i}{q_i}\Bigg).
\end{equation*}

\end{itemize}







\subsection*{Supplementary Text}

\subsection*{Global spatial connectivity using Reanalysis data}
\begin{figure}
\centering
\includegraphics[width=0.95\textwidth]{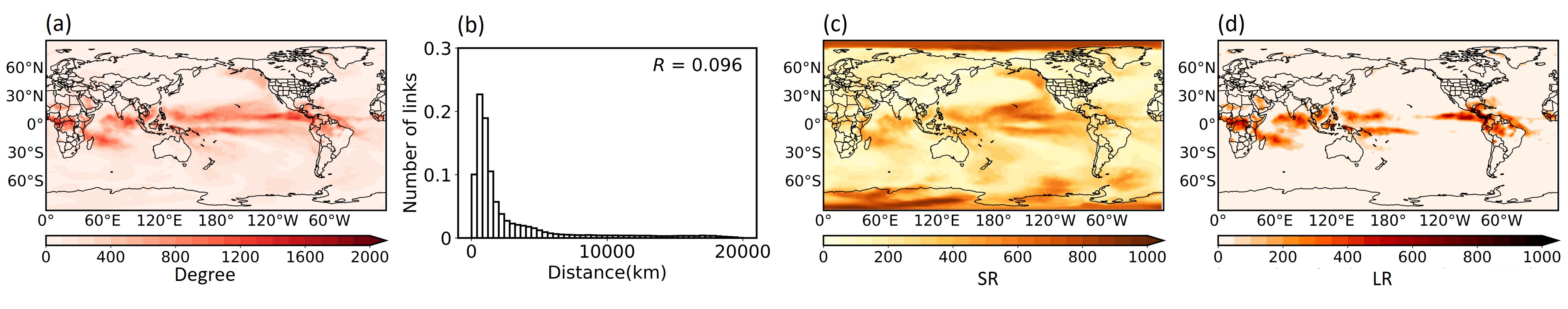}
\caption{\textbf{Properties of climate network generated from ERA5 reanalysis data ($1990$-$2019$).} We display (a) a plot of degree at the global spatial scale, (b) probability density plots of link lengths with the value of $R$, (c) the number of spatial short-range connectivity (`SR'), and (d) spatial long-range connectivity (`LR').}
\label{fig3} 
\end{figure}

\par Here, we investigate the spatial dependency structures derived from different datasets from the past: the ERA5 reanalysis dataset ($1990$-$2019$). The plots of weighted degree (Eq.\ \ref{eq.1}) associated with each node represent the number of connections for each geographical location based on reanalysis data (Fig.\ \ref{fig3}(a)). The normalized frequency of link lengths with connectivity ratio ($R$) (Eq.\ \ref{eq.3}) associated with ERA5 dataset gives an insight into the number of connection lengths against geographical distances which is another way to represent connectivity structure (Fig.\ \ref{fig3}(b)). 
We use two quantifiers that categorize the links for our purpose of the study: `SR' and `LR' of each node. The `SR' value of a node indicates the count of links that have lengths shorter than $5000$ km and the number of links exceeding a length of $10000$ km is designated as the `LR' value of a node. 
These maps of the numerical values of `SR' and `LR' for each node vividly demonstrate the differences in the number of long-range and short-range connections determined for ERA5 air temperature data and by selecting three distinct initial states of ESM (Figs.\ \ref{fig3}(c)-(d)) at the global scale.
Here, we demonstrate the result of ERA5 reanalysis data. We use this for comparison with future projections generated from EC-Earth3 and MPI-M ESMs.

\subsection*{Global connectivity structure for few selected regions}

\begin{figure*}
\centering
\includegraphics[scale=0.8]{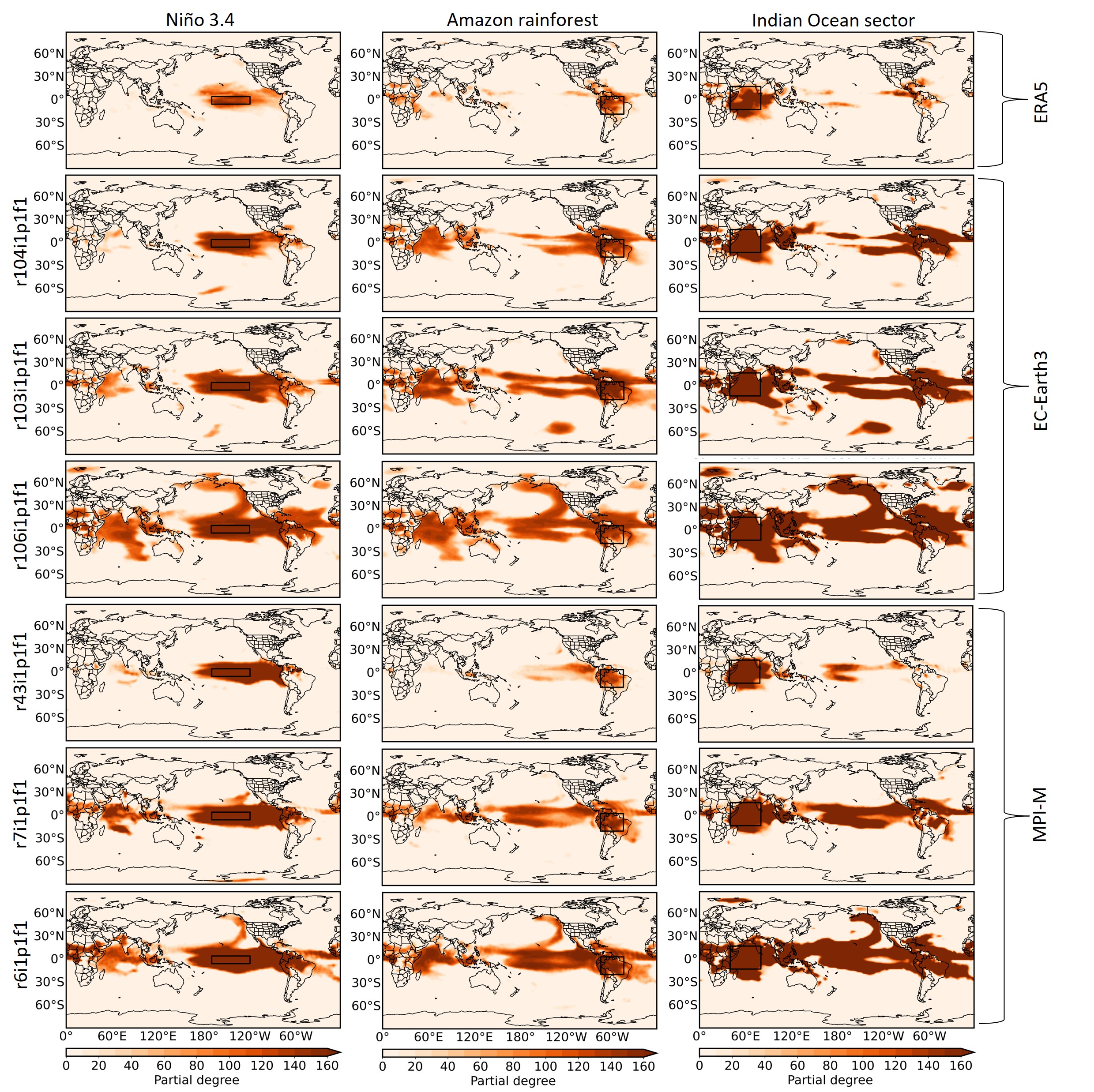}
\caption{\textbf{Partial degree of climate network for region-based analysis.} We draw the partial degree to analyze the connection pattern between a particular region and the entire planet. This study includes reanalysis data in the first row and model projections with three distinct initial conditions in the second, third, and fourth rows for EC-Earth3 ESM and the fifth, sixth, and seventh rows for MPI-M ESM, respectively.
The chosen three regions (shown in black boxes) are Ni{\~n}o 3.4 (left panel), the Amazon rainforest (middle panel), and a sector of the Indian Ocean (right panel), respectively. The outcomes for three different initial conditions are very distinguishable for the three regions.}
\label{fig4} 
\end{figure*}

\par We are analyzing the impact of ICV on the spatial connections from a particular region to the whole world by utilizing the concept of partial degree (Eq.\ \ref{eq.2}) in the climate network generated for air temperature reanalysis data of ERA 5 and three datasets for future projection of EC-Earth3 ESM and three datasets for future projection of MPI-M ESM (mentioned in the main text). 
We have selected three regions: (a) Ni{\~n}o $3.4$ ($5$\textdegree N-$5$\textdegree S, $120$\textdegree W-$170$\textdegree W), (b) the Amazon rainforest ($5$\textdegree N-$18$\textdegree S, $45$\textdegree W-$75$\textdegree W), (c) the sector of the Indian Ocean ($18$\textdegree N-$12$\textdegree S, $40$\textdegree E-$80$\textdegree E). We choose Ni{\~n}o $3.4$ because the El Ni{\~n}o–Southern Oscillation is a dominant mode that drives the inherent fluctuations in climate \cite{ng2021impacts}. We select the Amazon rainforest because it is a prominent tipping element, teleconnected to other regions of the world  \cite{liu2023teleconnections}. We pick a sector of the Indian Ocean because recent studies show rapid warming in the Indian Ocean \cite{wenegrat2022century}. The plots of partial degree (Eq.\ \ref{eq.2}) for each node of the global temperature network for three selected regions exhibit significant variations in partial degree when the future projections of temperature are simulated for different initial conditions (Fig.\ \ref{fig4}). Our study focuses on comparing the spatial connections from the Ni{\~n}o 3.4 region, the Amazon rainforest, and a sector of the Indian Ocean to other regions across the globe, based on the choice of initial conditions in the ESMs. These regions play a crucial role in global climate dynamics, and the amplification of their interactions may drive shifts in weather patterns, including temperature anomalies, precipitation changes, and atmospheric circulation. These regions serve as critical hubs in the global climate system, and their interactions are vital in modulating atmospheric and oceanic processes. 
\par The number of partial degrees varies from low to high across the ensemble members, resulting in significantly different spatial connectivity patterns. This illustrates how small perturbations in the initial conditions can be magnified over time, causing substantial differences in spatial structures. We present the results of three selected initial conditions of two ESMs, including one of each model that exhibits a connectivity structure close to ERA 5 reanalysis data and another that is substantially different from it (Fig.\ \ref{fig4}) to portray the full range of possibilities in the spatial connectivity structure of surface temperature. These demonstrate that while one set of initial conditions can project a future that is comparable to our observations, another set can present an extreme possibility where the number of connections is qualitatively higher, leading to a dense connectivity pattern. Such cases may require our attention as the projected climate could lie beyond the range of possibilities that are typically generated from the multi-model ensembles. For example, under present conditions, connectivity from the selected study regions to the tropical North Atlantic \cite{casselman2023teleconnection} and Pacific Meridional Mode \cite{fan2021influence, zhang2017impacts} has strong impacts on regional climate. While some initial conditions exhibit a connectivity structure that is close to the present conditions, some initial conditions show an enhanced connectivity structure in the region that could have a strong impact on the regional climate, particularly in the western part of North America. The variability of inter-regional connectivity over large ensemble simulations highlights the importance of understanding these dynamic relationships to better predict and mitigate the impacts of climate change, especially in vulnerable regions that are already experiencing significant climate-related challenges. 
  
\par The variations in the spatial connectivity structure at the regional scale across different initial conditions can be illustrated using the frequency plots of $R$ (Fig.\ \ref{fig7}). $R$$=$$1$ separates from long-range dominance over short-range dominance ($R > 1$) to short-range dominance compare to long-range dominance  ($R < 1$) across initial conditions. We compare the number of initial conditions where $R$ exceeds $1$ to the number of initial conditions where $R$ does not exceed $1$ for EC-Earth3 and MPI-M ESMs across three regions and two climate time scales (Table\ \ref{tab:model_comparison}). For the EC-Earth3 ESM, the number of initial conditions showing long-range dominance over short-range connections decreases in the long-term climate scale compared to the near-term scale for the Niño 3.4 and Amazon rainforest regions, while it remains unchanged for the Indian Ocean sector. For the Indian Ocean sector, $100\%$ initial conditions agree for the dominance of long-range spatial connections over short-range spatial connections for both near-term and long-term climate scales. But, the change of variations in the frequency plots of $R$ (Fig.\ \ref{fig7}) confirms the change of predictability in between two climate scales. For the MPI-M ESM, the number of initial conditions exhibiting long-range dominance over short-range connections increases in the long-term climate scale compared to the near-term scale for the Niño 3.4 region, but decreases for the Amazon rainforest and Indian Ocean sectors.

\begin{table}[h!]
\centering
\renewcommand{\arraystretch}{1.5}
\begin{tabular}{|c|c|c|c|c|c|c|c|c|}
\hline
\multicolumn{1}{|c|}{} &
\multicolumn{4}{c|}{\textbf{EC-Earth3}} &
\multicolumn{4}{c|}{\textbf{MPI-M}} \\
\cline{2-9}
 & \multicolumn{2}{c|}{\textbf{2020-2049}} & \multicolumn{2}{c|}{\textbf{2070-2099}} & \multicolumn{2}{c|}{\textbf{2020-2049}} & \multicolumn{2}{c|}{\textbf{2070-2099}} \\
\cline{2-9}
 & R$<$1 & R$>$1 & R$<$1 & R$>$1 & R$<$1 & R$>$1 & R$<$1 & R$>$1 \\
\hline
\textbf{Niño 3.4} & 66.67\% & 33.33\% & 100\% & 0\% & 62\% & 38\% & 42\% & 58\% \\
\hline
\textbf{Amazon rainforest} & 12.5\% & 87.5\% & 33.33\% & 66.67\% & 40\% & 60\% & 62\% & 38\%\\
\hline
\textbf{Sector of Indian Ocean} & 0\% & 100\% & 0\% & 100\% & 10\% & 90\% & 20\% & 80\%\\
\hline
\end{tabular}
\caption{Comparison of the number of initial conditions where $R$ either exceeds $1$ or does not exceed $1$ for the EC-Earth3 and MPI-M models, analyzed across different regions and climate timescales.}
\label{tab:model_comparison}
\end{table}

\subsection*{Visualization of spatial connectivity for large ensemble with quantifier of internal climate variability}
\par
\begin{figure*}
\centerline{
\includegraphics[scale=0.36]{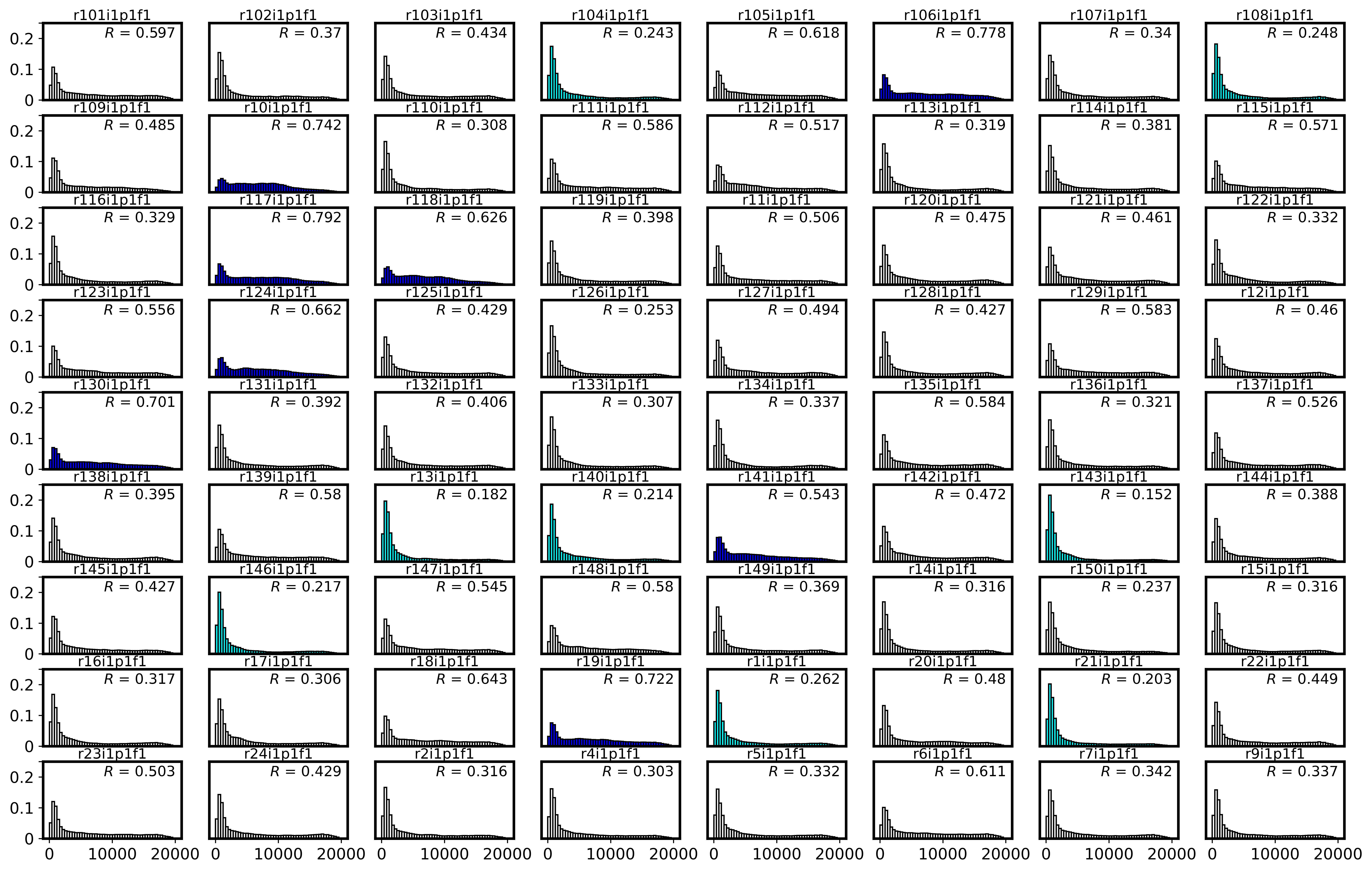}}
\caption{{\bf The normalized frequency plots of the link lengths (distance in km) for EC-Earth3 ESM}.  We display these for the $72$ climate networks generated from the $72$ plausible outcomes of future projections from EC-Earth3 ESM. The initial conditions for each plot are indicated at the top of the plots. Observations of the histogram plots highlight fluctuations in the spatial topology of global temperature connectivity. The tails of these distributions signify the presence of long-range spatial connections. Networks with visually heavier tails in the density plots exhibit a higher number of connections compared to others. The density plots with distances (Jensen-Shannon divergence) from the density plot obtained using ERA5 reanalysis data falling below the $10$-th percentile of all $72$ distances are highlighted in cyan. Conversely, those with distances exceeding the $90$-th percentile are marked in blue. The value of $R$ in each histogram denotes the ratio of long-range connections to short-range connections.}
\label{fig5} 
\end{figure*}

\begin{figure*}
\centerline{
\includegraphics[scale=0.36]{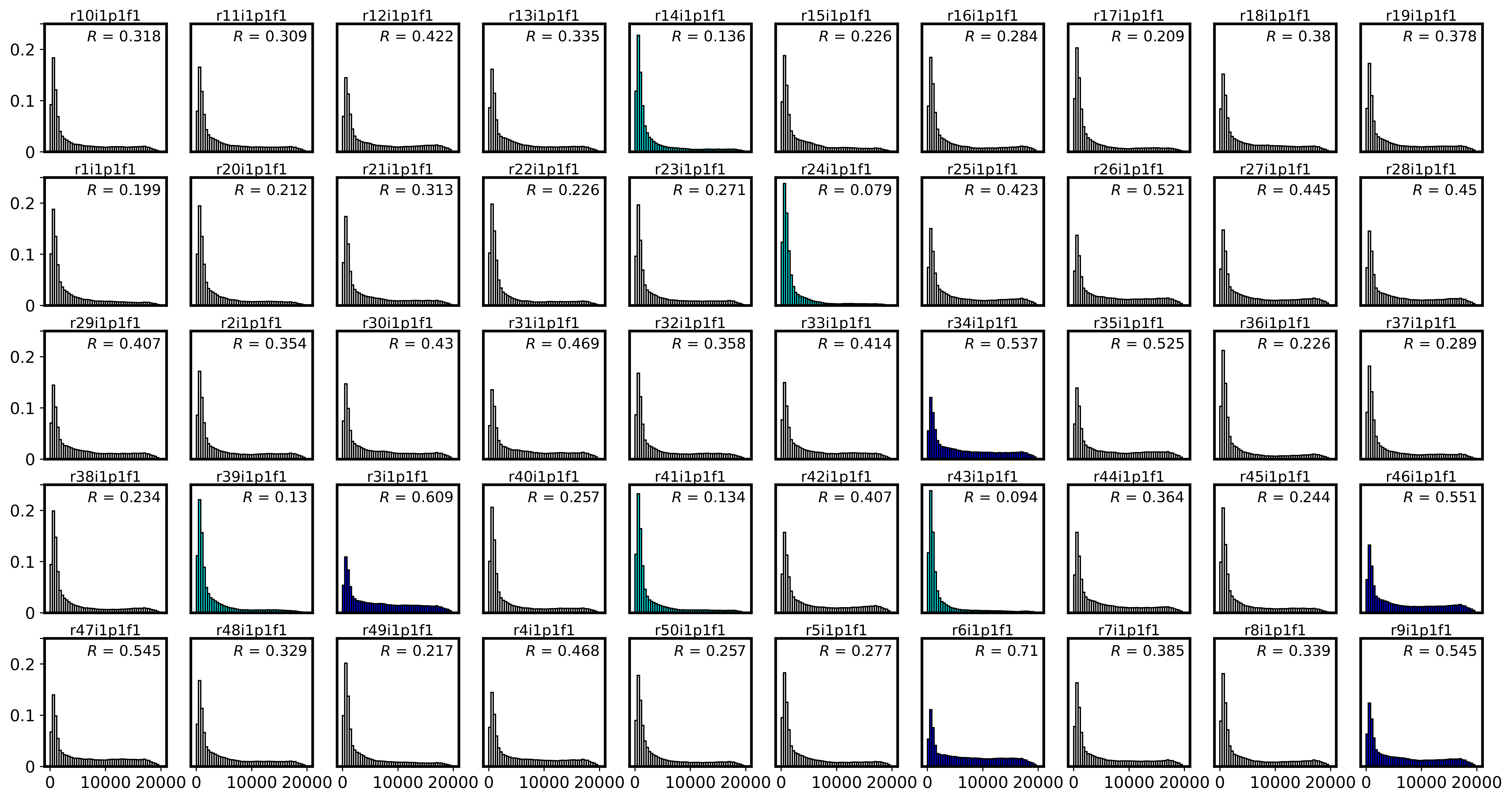}}
\caption{{\bf The normalized frequency plots of the link lengths (distance in km) for MPI-M ESM}.  We display these for the $50$ climate networks generated from the $50$ plausible outcomes of future projections from MPI-M ESM. The description is the same as Fig.\ \ref{fig5}.}
\label{fig6} 
\end{figure*}

 For visualization of spatial connectivity structure corresponding to each dataset, we draw normalized frequency plots of link lengths with quantifier, $R$ for $72$ initial conditions of the EC-Earth3 ESM (Fig.\ \ref{fig5}) and for $50$ initial conditions of the MPI-M ESM (Fig.\ \ref{fig6}). Both figures exhibit the variation across the initial conditions. Now, the question is whether we can categorize similar structures. For this, we employ the Jensen-Shannon divergence, which is used to calculate the difference between two probability density functions (Eq.\ \ref{eq.4}). We now compute the distance between two probability density functions of link lengths: one is derived using ERA5 reanalysis data (considered reference data), while the other corresponds to each initial condition obtained from two ESMs. The derived dataset of the distances enables us to characterize the structure of climate networks. Here, the $10$-th percentile and the $90$-th percentile of two datasets of distances are calculated to capture the datasets whose connectivity structure is close and extremely different from the reference dataset.  We point out the distributions that are situated at a distance less than the $10$-th percentile (cyan in Fig.\ \ref{fig5} Fig.\ \ref{fig6}) and those that exceed the $90$-th percentile (blue in Fig.\ \ref{fig5} and Fig.\ \ref{fig6}) from the distribution obtained using ERA5 data, presenting them in blue. The initial conditions whose corresponding density plots of link lengths are cyan, the spatial connectivity structures generated by those are mostly close to the structure created from ERA5 reanalysis data and less number of long-range connections out of $72$ initial conditions for EC-Earth3 ESM and $50$ initial conditions for MPI-M ESM. The initial conditions with blue density plots of link lengths reveal those spatial connectivity architectures that are significantly different from the structure formed by ERA5 reanalysis data and have a higher number of long-range connections. 

\end{document}